\documentclass[twocolumn,superscriptaddress,preprintnumbers,amsmath,10pt,aps,prl]{revtex4}
\usepackage{graphicx}
\usepackage{amsmath}
\usepackage{amsfonts}
\usepackage{bm}
\usepackage{color}
\begin{document}

\title{Imaging high-dimensional spatial entanglement with a camera}

\author{M.~P.~Edgar}
\email{matthew.edgar@glasgow.ac.uk}
\affiliation{SUPA, School of Physics and Astronomy, University of Glasgow, Glasgow, G12 8QQ, UK}
\author{D.~S.~Tasca}
\affiliation{SUPA, School of Physics and Astronomy, University of Glasgow, Glasgow, G12 8QQ, UK}
\author{F.~Izdebski}
\affiliation{SUPA, School of Engineering and Physical Sciences, Heriot-Watt University, Edinburgh, EH14 4AS, UK}
\author{R.~E.~Warburton}
\affiliation{SUPA, School of Engineering and Physical Sciences, Heriot-Watt University, Edinburgh, EH14 4AS, UK}
\author{J.~Leach} 
\affiliation{Department of Physics, University of Ottawa, Ottawa, ON K1N 6N5, Canada}
\author{M.~Agnew} 
\affiliation{Department of Physics, University of Ottawa, Ottawa, ON K1N 6N5, Canada}
\author{G.~S.~Buller} 
\affiliation{SUPA, School of Engineering and Physical Sciences, Heriot-Watt University, Edinburgh, EH14 4AS, UK}
\author{R.~W.~Boyd} 
\affiliation{Department of Physics, University of Ottawa, Ottawa, ON K1N 6N5, Canada}
\author{M.~J.~Padgett}
\affiliation{SUPA, School of Physics and Astronomy, University of Glasgow, Glasgow, G12 8QQ, UK}

\maketitle

The light produced by parametric down-conversion shows strong spatial entanglement that leads to violations of EPR criteria for separability. Historically, such studies have been performed by scanning a single-element, single-photon detector across a detection plane. Here we show that modern electron-multiplying CCD cameras can measure correlations in both position and momentum across a multi-pixel field of view. This capability allows us to observe entanglement of around 2500 spatial states and demonstrate Einstein-Podolsky-Rosen (EPR) type correlations by more than two orders of magnitude. More generally, our work shows that cameras can lead to important new capabilities in quantum optics and quantum information science.

Over the last two decades the spatial correlations between photon pairs produced by spontaneous parametric down conversion (SPDC) have been investigated in a number of different configurations, playing an important role in fundamental tests of quantum mechanics and in applications for quantum communication and information processing \cite{Walborn2010}. The high dimensionality of the transverse spatial degrees of freedom (DOF) of the photon pairs can be explored either by projections onto a discrete basis \cite{Dada2011}, such as the Laguerre-Gaussian modes, or by using a continuous basis defined by the transverse position or momentum of the photons \cite{Tasca2011}. For instance, it has been shown that their transverse position and momentum display EPR-type correlations \cite{Howell2004,Dangelo2004}. However, most measurements to date have relied upon the scanning of a single avalanche photodiode or a very small number of individual detectors \cite{Leach2012}. This sequential scanning or use of a small number of detectors negates any information capacity advantage in the use of spatial states. In all protocols for high-dimensional quantum key distribution \cite{Walborn2006,Walborn2008}, quantum computation \cite{Tasca2011} and teleportation \cite{Walborn2007} using spatial states, it is essential to perform a full field measurement of the photon transverse position, which is made possible by using a 2D detector array.

As with any quantum observable, the transverse position or momentum of single photons can only be measured with a limited resolution \cite{Rudnicki2011}. In most of the cases, spatial entanglement has been detected through measurements of intensity correlation by scanning the detectors in two conjugate planes \cite{Howell2004,Tasca2008,Tasca2009,Gomes2009,Walborn2011}. By knowing the intensity correlation distributions, non-separability \cite{Tasca2008,Gomes2009} or EPR \cite{Howell2004,Tasca2009,Walborn2011} criteria could be tested. In these cases, the resolution of the measurements is limited by the size of the pinhole or slit used for the scanning. In Ref. \cite{Pires2009}, the authors performed a direct measurement of the Schmidt number associated with the transverse distribution of the two-photon field, where the intensity distribution of only one of the down-converted fields was measured in both image and far field planes with the help of an intensified CCD. In this case, the resolution of the measurement is limited by the size of the camera pixels. Sub-shot-noise measurements of correlations in intensity fluctuations in SPDC have already been performed with the help of a CCD camera \cite{Jedrkiewicz2004}. In another experiment a CCD camera was used  to measure the complementarity between one and two-photon interference from a Young's double slit \cite{Sasha2001}. Nevertheless, noise in CCD cameras prevents single photon discrimination.

Recent years have seen a rapid advance in imaging technologies. Developments in low-noise electron multiplying CCD (EMCCD) cameras suggest they are capable of greater than $90\%$ quantum efficiencies and, if operated in a low light environment whilst being maintained at low temperatures, can provide single photon sensitivity \cite{Lantz2008}. EMCCDs have been used in measurements of sub-shot-noise correlations of intensity fluctuations \cite{Blanchet2008,Brida2010} as well as in intensity correlations \cite{Zhang2009} in the far-field of the photon pairs from SPDC.

Here we report the observation of spatial correlations from SPDC using an EMCCD camera in both the image plane and far-field of the non-linear crystal showing position correlation and momentum anti-correlation, respectively. In both cases the images are sparse, that is the number of recorded photons per pixel is around $0.02$. We make no post selection of optimum images \cite{Oemrawsingh2002}; rather we obtain correlations by summing over all the recorded frames. The use of a 2D array of detectors allows the exploration of both spatial DOF of the photons at the same time, thereby increasing the number of accessible states and the information content per photon \cite{Dixon2011}. Here we use a $201 \times 201$ array of pixels, to show 2D correlations in both position and momentum of the photon pairs. In both bases we show correlations of around 50 modes in both transverse directions, giving in principle access to several thousand entangled spatial states.

\section{Results}
\begin{figure}[b]
\begin{center}
\includegraphics[width=3.3in]{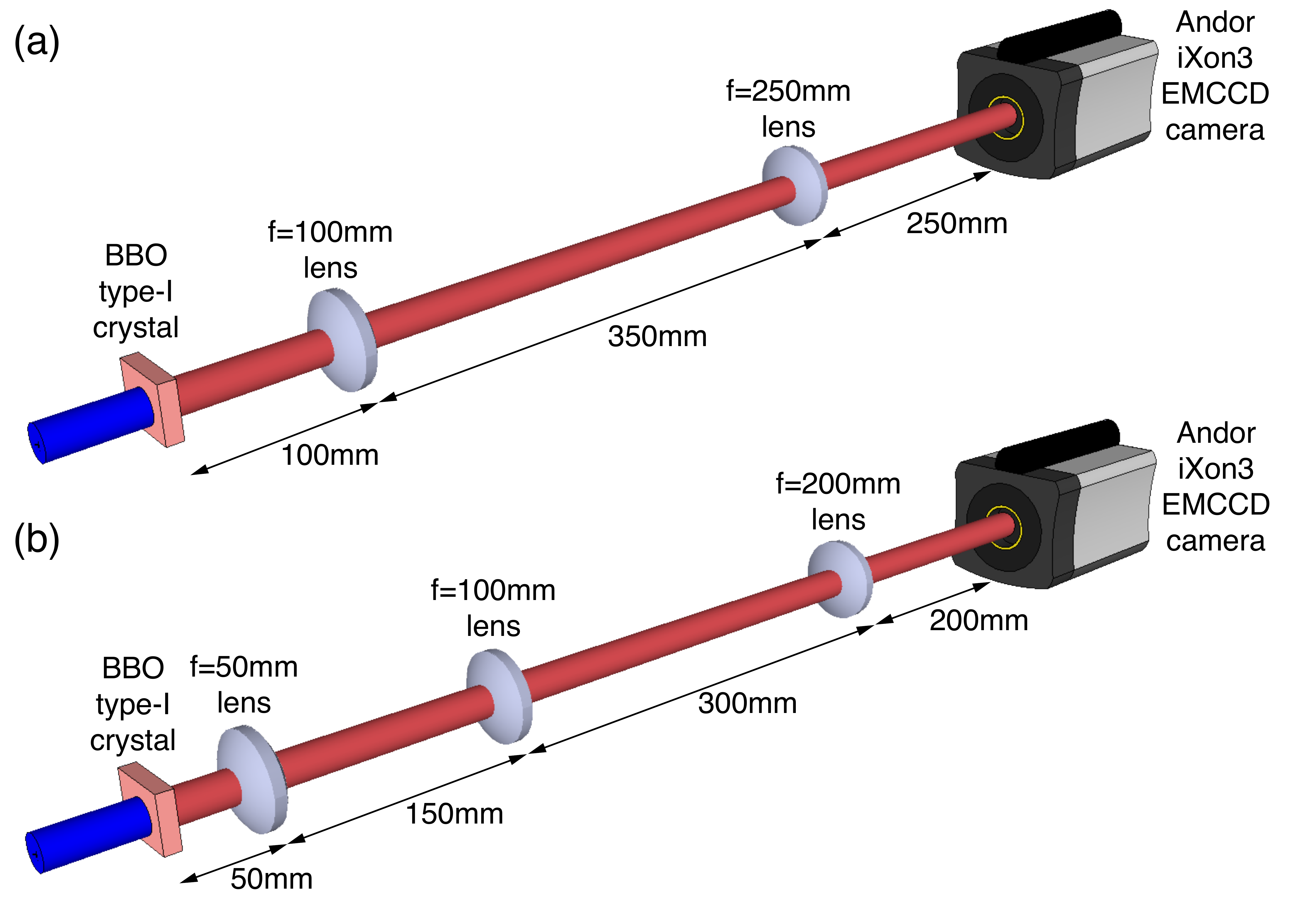}
\caption{{\bf Experimental scheme used to measure position and momentum correlations.} ${\rm ({\bf a})}$ The imaging system used to measure position correlations consists of a telescope with lenses $100\,\rm{mm}$ and $250\,\rm{mm}$. ${\rm ({\bf b})}$ Momentum correlations are obtained with the implementation of three consecutive Fourier systems with lenses $50\,\rm{mm}$, $100\,\rm{mm}$ and $200\,\rm{mm}$.}
\label{Experimental_Setup}
\end{center}
\end{figure}

{\bf Imaging system.} The experimental setup used to measure spatial correlations in both position and momentum is shown in Fig.\,\ref{Experimental_Setup}. A $150\,\rm{mW}$, high-repetition-rate $355\,\rm{nm}$ laser is attenuated to $2\,\rm{mW}$ and using a simple telescope the Gaussian beam radius is expanded to $\sigma_{p} = (0.66\pm0.05)\rm{mm}$. A $5\,\rm{mm}$ long $\beta$-Barium Borate (BBO) crystal cut for type-I phase matching was angled to provide near-collinear output for degenerate down-converted photons at $710\,{\rm nm}$. The subsequent choice of lens configurations was chosen to maintain a constant separation ($700\,\rm{mm}$) between the BBO crystal and the EMCCD (Andor iXon3), thus providing an efficient system for manual switching between the image and far-field planes. 

As used by other authors \cite{Law2004,Tasca2009}, we use a Gaussian model to describe the spatial structure of the two photon field at the crystal:
\begin{equation}
\Psi(\boldsymbol{\rho}_1,\boldsymbol{\rho}_2) = N\exp\left[-\frac{|\boldsymbol{\rho}_1+\boldsymbol{\rho}_2|^2}{4\sigma_+^2}\right]\exp\left[-\frac{|\boldsymbol{\rho}_1-\boldsymbol{\rho}_2|^2}{4\sigma_-^2}\right],
\label{spatialstructure}
\end{equation}
where $N = 1/(\pi\sigma_-\sigma_+)$ is a normalization constant, $\boldsymbol{\rho}_i = (x_i,y_i)$ is the transverse position of photon $i$ ($i = 1,2$) and $\sigma_{\pm}$ are the standard deviations of the two Gaussians, giving the strength of the position and momentum correlations, $\sigma_-$ and $\sigma_+^{-1}$, respectively. We define $\sigma_+$ to be equal to the standard deviation of the Gaussian field distribution of the pump laser, $\sigma_{p}$, with wavelength $\lambda_{p}$ and 
\begin{equation}
\sigma_- = \sqrt{\frac{\alpha L\lambda_{p}}{2\pi}},
\label{standarddevpos}
\end{equation}
where $L$ is the length of the SPDC crystal and $\alpha = 0.455$ is an adjustment constant \cite{Chan2007}. Eq.\,\eqref{spatialstructure} is referred to as the transverse wave function of the post-selected two-photon field of SPDC for a Gaussian pump beam \cite{Monken98a}. The joint detection probability is $\mathcal{P}(\boldsymbol{\rho}_1,\boldsymbol{\rho}_2)\propto|\Psi(\boldsymbol{\rho}_1,\boldsymbol{\rho}_2)|^2$. According to the Gaussian model and the parameters of our system we predict the number of modes for joint detections in both position and momentum to be $\left(\sigma_+/\sigma_-\right)^2 \approx 3500$.

\begin{figure}[t]
\begin{center}
\includegraphics[width=3.3in]{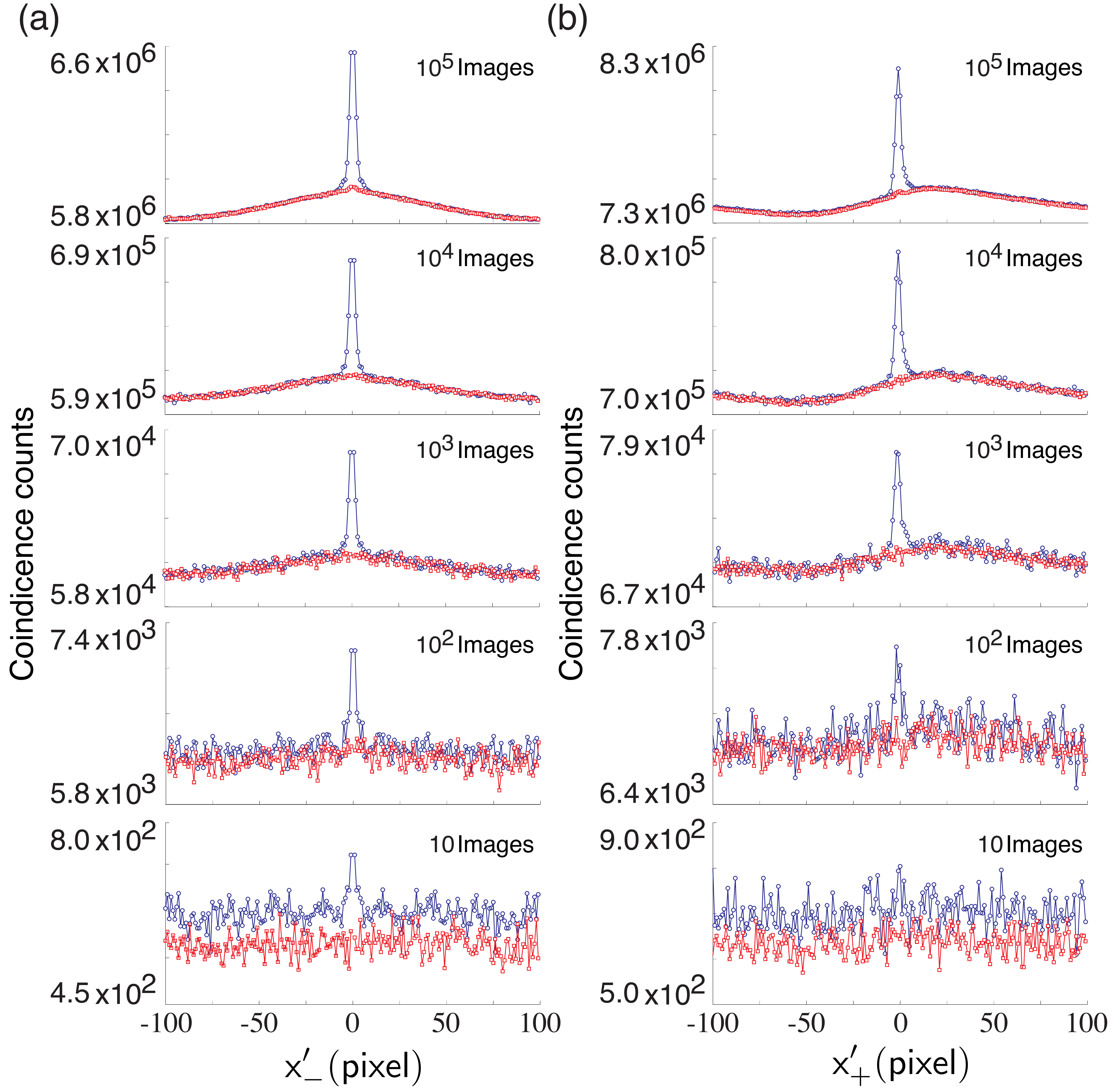}
\caption{{\bf Position and momentum correlation functions in the $x'$ dimension for an increasing number of images.} ${\rm ({\bf a})}$ Measured position correlations from imaging plane scheme. ${\rm ({\bf b})}$ Measured momentum correlations from far-field scheme. Blue circles represents the spatial-correlation and red squares represent the reference-correlation. The asymmetry of the far-field correlation function is the result of a non-uniform near collinear degenerate light cone.}
\label{Mom_Pos_Cross}
\end{center}
\end{figure}

When measuring position correlations, we use an imaging system with a magnification of $M = 2.5$, comprising two lenses of focal lengths $100\,\rm{mm}$ and $250\,\rm{mm}$ (Fig.\,\ref{Experimental_Setup}a). The far-field image is obtained by using a composite Fourier system with an effective focal length of $f_{e} = 100\,{\rm mm}$ (Fig.\,\ref{Experimental_Setup}b). Given our imaging and far-field system we predict a transverse correlation length $\sigma_{{\rm pos}} = M\sigma_- \approx 28\,\mu\rm{m}$ and $\sigma_{{\rm mom}} = f_e/(k\sigma_+) \approx17\,\mu\rm{m}$, where $k = k_{p}/2$ is the wavenumber of the down-converted photons. Note that the position correlation length is larger than the pixel size of the camera ($16\,\mu\rm{m}$) meaning that the position correlated photon pairs have a high probability of being detected in adjacent pixels which removes the need to count multiple photons within the same pixel. 

A $10\,\rm{nm}$ bandpass filter centred at $710\,\rm{nm}$ with a transmission efficiency $\eta_{{\rm filter}} = 90\%$ is located immediately in front of the EMCCD. The EMCCD camera used in our experiment is a back-illuminated $512\times512$ array of $16\times16\,\mu\rm{m^2}$ pixels optimized for visible wavelengths. The quantum efficiency of detection is given as the product of the efficiency of the camera, filter and lenses. Thus providing that either the signal or idler photon is detected, we predicted a heralding efficiency for detecting its entangled partner to be $\eta \approx 80\%$.

{\bf Image analysis.} A series of $100,000$ images was recorded for image and far-field configurations (see Methods). The probability distributions $\mathcal{P}(\boldsymbol{\rho}_-)$ and $\mathcal{P}(\boldsymbol{p}_+)$ were calculated respectively by counting the number of coincidences as a function of the difference (imaging plane) or sum (far-field) of the pixel coordinates in both $x'$ and $y'$ directions at the plane of the CCD array. Here we define $\boldsymbol{\rho}_- \equiv \boldsymbol{\rho}_1-\boldsymbol{\rho}_2$ and $\boldsymbol{p}_+ \equiv \boldsymbol{p}_1 + \boldsymbol{p}_2$, where $\boldsymbol{p}_i = (p_{x_i},p_{y_i})$ is the transverse momentum of photon $i$. The transformation from coordinates in the plane of the CCD array to the crystal is performed using the scaling factors $\gamma_{\rm{pos}} \equiv 1/M$ for imaging plane and $\gamma_{\rm{mom}} \equiv (k \hbar)/f_{e}$ for far-field. Our result is obtained by averaging the spatial correlation function over all recorded frames. Within each linear dimension this processing reveals correlations between photon pairs, which are clearly visible on top of a broader correlation arising from uncorrelated events, as shown in Fig.\,\ref{Mom_Pos_Cross}. We measured the background correlation by performing the same analysis but between consecutive frames to give a reference correlation. Fig.\,\ref{Mom_Pos_Cross}a shows the measured coincidence counts in the image plane as a function of $x'_{-}$ coordinates, while coincidence counts measured in the far-field as a function of $x'_{+}$ are shown in Fig.\,\ref{Mom_Pos_Cross}b. A correlation peak in both position and momentum can already be seen in a few images and becomes significantly distinguishable above noise as more images are summed.
\begin{figure}[b]
\begin{center}
\includegraphics[width=3.3in]{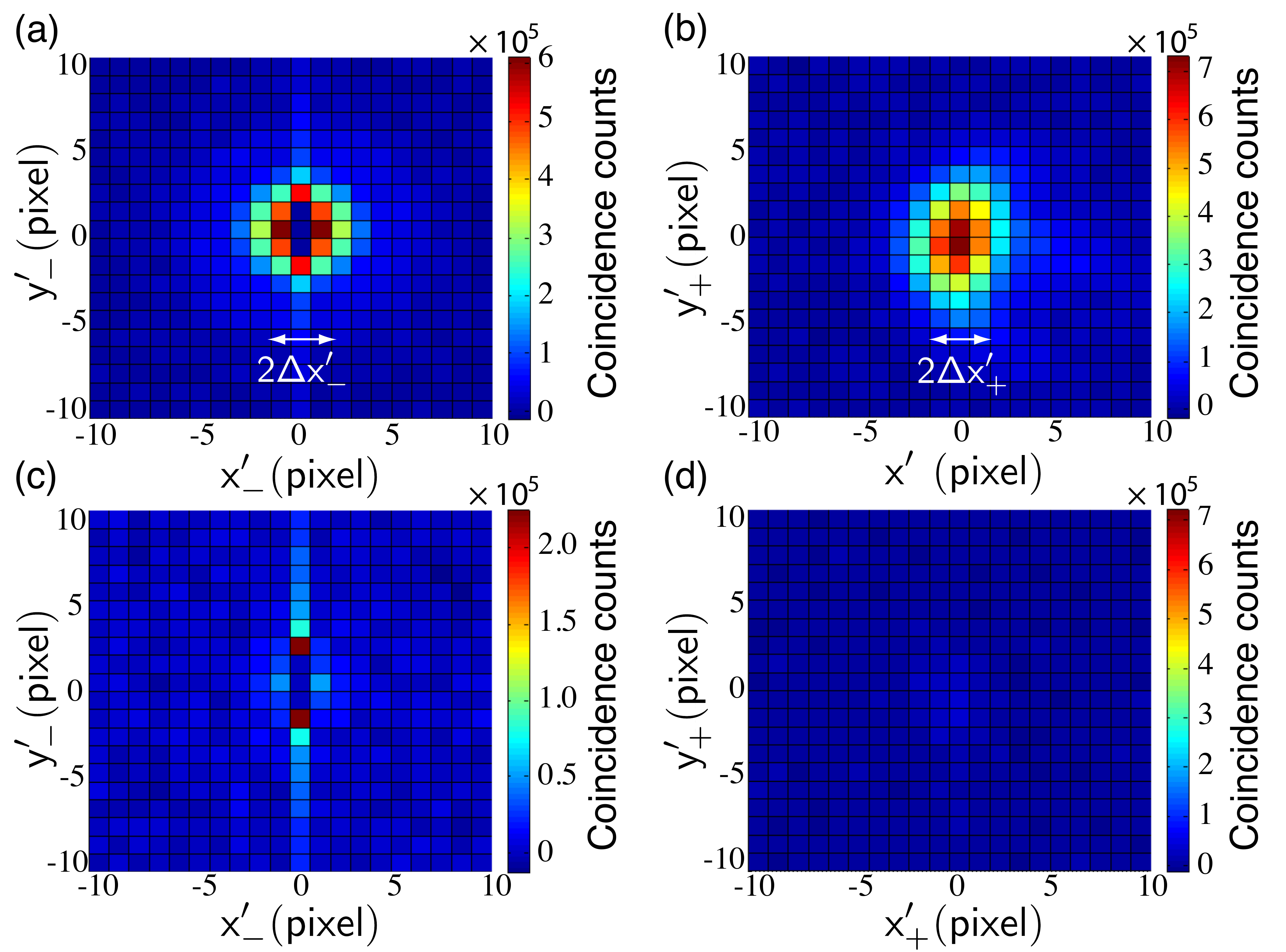}
\caption{{\bf Background subtracted auto-correlation functions in position and momentum for high and low heralding efficiency configurations.} Auto-correlation functions for the image plane ${\rm ({\bf a})}$ and far-field ${\rm ({\bf b})}$ of the crystal, summed over 100 000 images and with the measured background correlation removed. In ${\rm ({\bf c})}$ and ${\rm ({\bf d})}$ we show the same measurements performed with a lower heralding efficiency. Variances along the $x'$-direction of the high heralding efficiency configuration are indicated. Due to increased coincidence counts attributed to smearing, the correlation function for $y'_{-} = \pm1$ were set to 0 in ${\rm (a)}$ and ${\rm (c)}$ for clarity.}
\label{Marginal_Correlations}
\end{center}
\end{figure}

By subtracting the reference from the spatial correlation function it is possible to remove unwanted pixel defects and obtain a strong correlation in position and anti-correlation in momentum as indicated in Fig.\,\ref{Marginal_Correlations}a and Fig.\,\ref{Marginal_Correlations}b respectively. Since we cannot distinguish the detection of more than one photon on the same pixel, we are unable to measure the correlation function for ${\boldsymbol \rho}'_1 = {\boldsymbol \rho}'_2$, for which a value equal to $0$ has been assigned. To confirm that these correlations arise between photon pairs the neutral density filter used to attenuate the pump beam was placed after the BBO crystal to reduce the heralding efficiency to approximately $2.0\%$, whilst ensuring the same photon flux. As shown in Fig.\,\ref{Marginal_Correlations}c and Fig.\,\ref{Marginal_Correlations}d, we do not observe a strong correlation in these measurements; however along the $y'_{-}$ axis in the image plane measurements we observe a correlation indicative of charge smearing between pixels in the readout direction.

\begin{figure}[t]
\begin{center}
\includegraphics[width=3.3in]{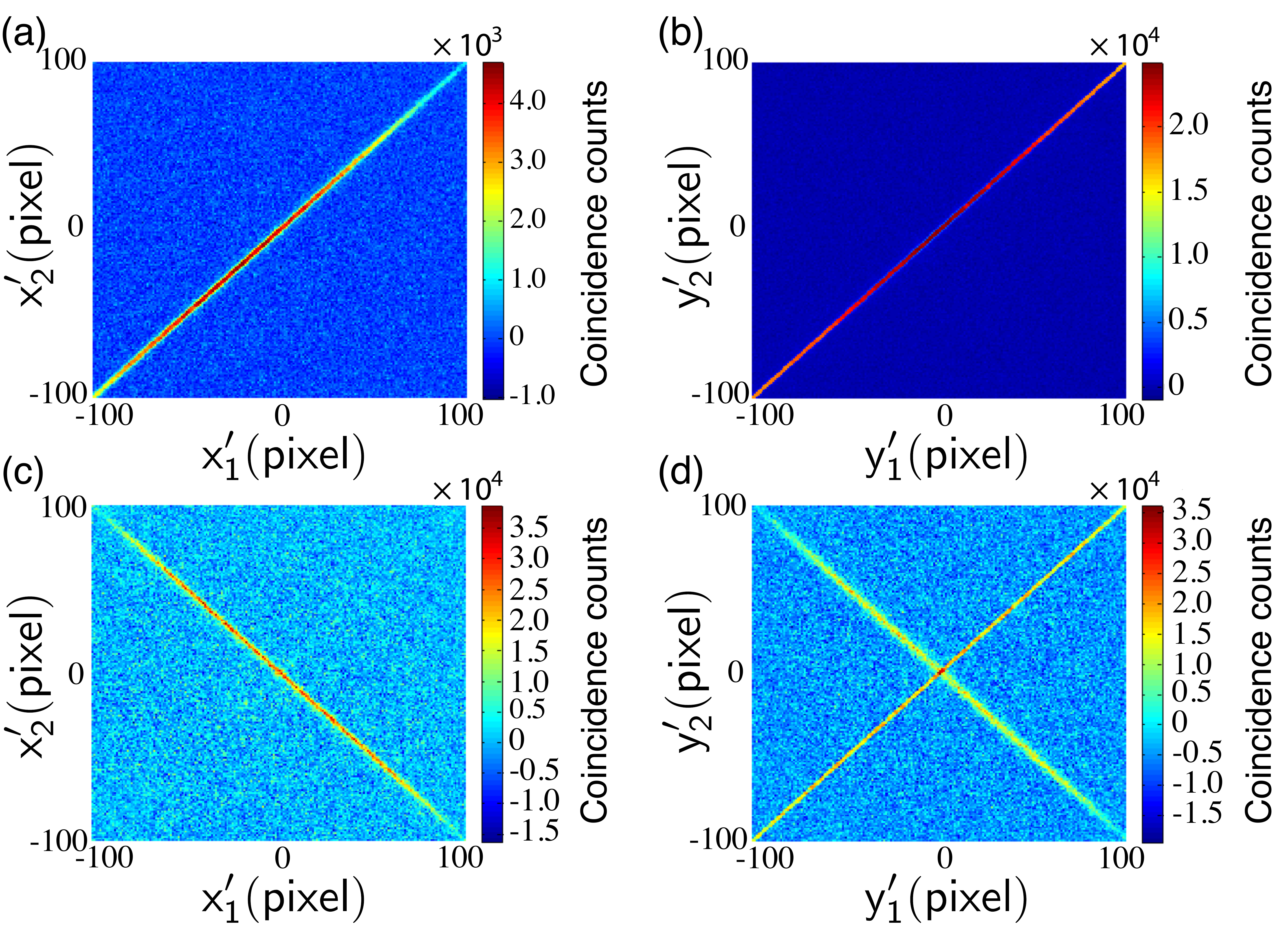}
\caption{{\bf Joint probability distributions for $x'$ and $y'$-coordinates.} The probability distributions for joint detections in both $x'$ and $y'$-coordinates in the image plane ${\rm ({\bf a}}$ and ${\rm {\bf b})}$ and far-field ${\rm ({\bf c}}$ and ${\rm {\bf d})}$ are shown respectively. Charge smearing gives rise to an artificial correlation in the $y'$ direction ${\rm ({\bf b}}$ and ${\rm {\bf d})}$.}
\label{Joint_Correlations}
\end{center}
\end{figure}

{\bf EPR-type correlations.} From the thresholded images we can calculate the joint probability distributions for both the $x'$ and $y'$ coordinates in the image plane and far-field, as shown in Fig.\,\ref{Joint_Correlations}. Fig.\,\ref{Joint_Correlations}a and Fig.\,\ref{Joint_Correlations}b show the coincidence counts in the planes $(x_1,x_2)$ and $(y_1,y_2)$, while Fig.\,\ref{Joint_Correlations}c and Fig.\,\ref{Joint_Correlations}d show the coincidence counts in the planes $(p_{x_1},p_{x_2})$ and $(p_{y_1},p_{y_2})$. In the $y'$ direction we observe a strong correlation resulting from charge smearing, thus preventing any meaningful analysis from image plane measurements (Fig.\,\ref{Joint_Correlations}b); however the expected anti-correlation in the far-field (Fig.\,\ref{Joint_Correlations}d) is still evident. Note that background subtraction has been performed.
It has been shown that EPR-like correlations can be identified by violating the inequality \cite{Reid1989,Reid2009}
\begin{equation}
\Delta^2_{\rm{min}}(x_1|x_2)\Delta^2_{\rm{min}}(p_{x_1}|p_{x_2})>\frac{\hbar^2}{4},
\label{EPRcorrelations}
\end{equation}
where $\Delta^2_{\rm{min}}(r_1|r_2)$ is the minimum inferred variance, describing the minimum uncertainty in measuring the variable $r_1$ conditional on the measurement of variable $r_2$. The minimum inferred variance for $x$-coordinates in the far-field measurement is defined as
\begin{equation}
\Delta^2_{\rm{min}}(p_{x_1}|p_{x_2})=\int \mathcal{P}(p_{x_2})\Delta^2(p_{x_1}|p_{x_2})dp_{x_2}.
\label{MinVarianceMomX1}
\end{equation}
We define the minimum inferred variance for the $x$-coordinates in the image plane with a similar approach. Fitting Gaussians to the joint probabilities in the $x$-direction (Fig.\,\ref{Joint_Correlations}a and Fig.\,\ref{Joint_Correlations}c) we obtain the variances shown in Table \ref{Table_variances}. Substituting these quantities into Eq.\,\eqref{EPRcorrelations} we find
\begin{eqnarray}
\Delta^2_{\rm{min}}(x_1|x_2)\Delta^2_{\rm{min}}(p_{x_1}|p_{x_2})=&(6.6\pm 1.0)\times10^{-4}\hbar^2,\\
\label{EPRcorrelationsX1}
\Delta^2_{\rm{min}}(x_2|x_1)\Delta^2_{\rm{min}}(p_{x_2}|p_{x_1})=&(6.2\pm0.9)\times10^{-4}\hbar^2,
\label{EPRcorrelationsX2}
\end{eqnarray}
indicating that the transverse DOF of the two photon field exhibits EPR non-locality. This result is in good agreement with the theoretical prediction (from Eqs.\,\eqref{spatialstructure} and
\eqref{MinVarianceMomX1}) of $\approx3\times10^{-4}\hbar^2$ and, although performed with background subtraction, is an order of magnitude smaller than those in Refs.\,\cite{Howell2004,Tasca2009}. 

{\bf High-dimensional entanglement.} From the joint distributions in Fig.\,\ref{Joint_Correlations},  we can estimate the ratio $(\sigma_+/\sigma_-)$ for both $x$ and $y$ transverse DOF, in the image plane (IP) and far-field (FF). We define the maximum number of joint detections for our position and momentum measurements to be $D^{\rm max}_{\rm pos}=[(\sigma_{x'_+}/\sigma_{x'_-})(\sigma_{y'_+}/\sigma_{y'_-})]_{\rm{IP}}$ and $D^{\rm max}_{\rm mom}=[(\sigma_{x'_-}/\sigma_{x'_+})(\sigma_{y'_-}/\sigma_{y'_+})]_{\rm{FF}}$. We must also account for the proportion of the beam measured by the detector array which is calculated by integrating the fitted Gaussian functions over the array size. In all cases we find this gives access to more than $85\%$ of the available states. Due to charge smearing in the image plane measurements we must estimate the dimensionality in $y$ based on that of the $x$ dimension, which is supported by the circular symmetry of Fig.\,\ref{Marginal_Correlations}a. Thus, the measured dimensionality in position and momentum is found to be $D_{\rm pos} = D^{\rm max}_{\rm pos} \times 0.95^2 \approx 3200$ and $D_{\rm mom} = D^{\rm max}_{\rm mom} \times 0.95 \times 0.85 \approx 2500$, respectively. To the best of our knowledge this represents the largest dimensionality for any experiment using entangled spatial states of photons.

\section{Discussion}
We have performed a multimode detection of around 2500 spatially entangled states of photon pairs produced by SPDC. We have shown that it is possible to utilize an EMCCD camera to measure spatial correlations between photon pairs in both the image plane and far-field of the down-conversion crystal. After background subtraction, we found that the spatial correlations violate EPR criterion. We acknowledge that the use of background-subtracted measurements do not meet the strict requirements for tests of EPR non-locality. If the background was included in the data, it would lead to an overestimate of the standard deviations of the correlations. Furthermore, in order to satisfy non-locality, the measurement of each photon must be performed outside the light-cone of its entangled partner. This issue could be solved in a modified setup in which the two-photons are split and sent to two spatially separated cameras. However, our results are in close agreement with the theoretical predictions, providing evidence that EMCCD technology can already be used in the characterization of spatial correlations of photons. 

The ability to access both transverse DOF of photons enables a large state space suggesting the use of spatial states in quantum communication and quantum information processing. In this case, information should be encoded and processed in an alphabet defined as orthogonal spatial states and subsequently read by projecting the full transverse field of the single photon on to a complete set of orthogonal spatial states. Thus, low-noise and efficient 2D detector arrays are essential in the use of spatial states in these applications, for which our work is a significant step forward. Future developments in imaging technologies will likely result in 2D detector arrays with even lower noise and widen their application in quantum optics.

\begin{table}[t]
\begin{center}\begin{tabular}{ccc}Variances & \, \, \, \, \, \, & Experimental Value\\ \hline \hline $\Delta^2_{\rm min}(x_1|x_2)$ & \, \, \, \, \, \, & $(119\pm 15){\rm \mu m}^2$ \\$\Delta^2_{\rm min}(x_2|x_1)$  & \, \, \, \, \, \, & $(119\pm 15){\rm \mu m}^2$ \\ $\Delta^2_{\rm min}(p_{x_1}|p_{x_2})$ & \, \, \, \, \, \, & $ (5.5\pm 0.4)\times10^{-6}\hbar^2{\rm \mu m}^{-2}$ \\ $\Delta^2_{\rm min}(p_{x_2}|p_{x_1})$ & \, \, \, \, \, \, & $(5.2\pm 0.4)\times10^{-6}\hbar^2{\rm \mu m}^{-2}$\\ \hline \end{tabular} \caption{Variances obtained from Gaussian fittings of background subtracted correlation functions shown Fig.\,\ref{Joint_Correlations}.}
 \label{Table_variances}
\end{center}
\end{table}

\section{Methods}

{\small {\bf Thresholding.} There are several noise sources that affect an EMCCD camera. As well as stray photons, the camera is subject to dark noise; thermally excited electrons, clock induced noise resulting from spurious electrons created during charge transfers between pixels, and readout noise generated by the readout register. We characterized the level of total noise in our camera when operated at the maximum gain by obtaining $100,000$ images while the laser is blocked but with the shutter on the camera open. Performing an analysis similar to that of Ref.\,\cite{Lantz2008} we find a Gaussian distribution of readout noise, centred at a value $390\,e^-$ with a standard deviation $\sigma_{\rm noise} = 6\,e^-$, in addition to a long tail that decreases exponentially for higher readout values. Our thresholding is performed by subtracting the mean background value for each pixel and assigning a $1$ to pixels with values greater than one standard deviation of noise, and a $0$ to all pixels having smaller values. This binary thresholding was found to reduce the level of noise whilst maximizing the correlation signal strength.}

{\small {\bf Camera Settings.} The camera was operated at $-85^0\rm{C}$, with a horizontal pixel shift readout rate of $1\,\rm{MHz}$, a vertical pixel shift every $0.3\,\mu\rm{s}$ and a vertical clock amplitude voltage of $+4$ above the default factory setting. The optimum photon flux incident upon the camera was chosen to maximize the strength of the correlation signal relative to noise sources. In our experiment the image-plane measurements were made with an exposure time of $0.4\,\rm{ms}$, while the far-field measurements are obtained using an exposure time of $1\,\rm{ms}$, in both cases giving an average photon flux of approximately 0.02 photons/pixel, which is equivalent to the level of noise. A region of interest measuring $201 \times201$ pixels was chosen to observe most of the down-converted field, which resulted in a frame rate of approximately $5\,{\rm Hz}$.}


{\bf Acknowledgements}

M.J.P. would like to thank the Royal Society and the Wolfson Foundation. M.J.P. and R.W.B. thank the US DARPA/DSO InPho program. M.P.E., D.S.T., F.I., R.E.W., G.S.B. and M.J.P. acknowledge the financial support from the UK EPSRC. J.L., M.A., and R.W.B thank Canada's CERC program. The authors would like to thank N.~Radwell for useful discussions.

{\bf Author contributions}

M.J.P., R.W.B., G.S.B. and J.L. conceived the experiment. M.P.E., D.S.T., F.I. and R.E.W. designed and performed the experiment. M.P.E., D.S.T., F.I., J.L., M.A. and M.J.P. designed the measurement and analysis programs. M.P.E., D.S.T., F.I., J.L. and M.J.P. analysed the results. All authors contributed to the writing of the manuscript. 

\end{document}